%% file: main.tex
\documentclass[preprint,12pt,authoryear]{elsarticle}

\usepackage{algorithmic}
\usepackage{tcolorbox}
\usepackage{textcomp}
\usepackage{xcolor}
\usepackage{hyperref}

\usepackage{graphicx}
\usepackage{natbib}
\usepackage{doi}
\usepackage{tikz}
\usepackage{graphicx,wrapfig,lipsum}
\usepackage{svg}
\usepackage{amsmath}
\usepackage{float}

\journal{ACM IMX'23 Interactive Media Experience}

\begin{document}

\begin{frontmatter}



\title{LLM-based Interaction for Content Generation: \\ A Case Study on the Perception of Employees in an IT department}


\author[label1,label2,label3]{Alexandre Agossah}

\affiliation[label1]{organization={Nantes Université, École Centrale Nantes, CNRS, LS2N, UMR 6004, F-44000 Nantes, France},%Department and Organization
           % addressline={Goodwin Hall}, 
            city={Nantes},
            country={France}}

\affiliation[label2]{organization={Digital Design Lab, L'École de design Nantes Atlantique},%Department and Organization
            %addressline={University of Manitoba}, 
            city={Nantes},
            country={France}}

\affiliation[label3]{organization={Groupe SIGMA},%Department and Organization
            %addressline={University of Manitoba}, 
            city={La Chapelle-sur-Erdre},
            country={France}}

\author[label2,label4]{Frédérique Krupa}

\affiliation[label4]{organization={Université Paris 1 Pantheon-Sorbonne},%Department and Organization
            %addressline={University of Manitoba}, 
            city={Paris},
            country={France}}
            
\author[label1]{Matthieu Perreira Da Silva}
\author[label1]{Patrick Le Callet}

\begin{abstract}
In the past years, AI has seen many advances in the field of NLP. This has led to the emergence of LLMs, such as the now famous GPT-3.5, which revolutionise the way humans can access or generate content. Current studies on LLM-based generative tools are mainly interested in the performance of such tools in generating relevant content (code, text or image). However, ethical concerns related to the design and use of generative tools seem to be growing, impacting the public acceptability for specific tasks. This paper presents a questionnaire survey to identify the intention to use generative tools by employees of an IT company in the context of their work. This survey is based on empirical models measuring intention to use (TAM by Davis, 1989, and UTAUT2 by Venkatesh and al., 2008). Our results indicate a rather average acceptability of generative tools, although the more useful the tool is perceived to be, the higher the intention to use seems to be. Furthermore, our analyses suggest that the frequency of use of generative tools is likely to be a key factor in understanding how employees perceive these tools in the context of their work. Following on from this work, we plan to investigate the nature of the requests that may be made to these tools by specific audiences.
\end{abstract}


% \begin{highlights}
% \item Perform an empirical study on the update request comments in Stack Overflow
% \item Provide an annotated dataset of 1,221 comments posted on 384 answers to Java questions
% \item Propose a supervised-learning approach to detect URCs with an average accuracy of 90\% 
% %\item Our approach can be used in Stack Overflow to decrease the rate of unaddressed URCs
% \end{highlights}


\end{frontmatter}



%%       \citet{<label>} ==> Jones et al. (2015)
%%       \citep{<label>} ==> (Jones et al., 2015)

\section{Introduction}\label{sec:introduction}
\input{introduction}

\section{The emergence of LLM-based tools for code, text and image generation}\label{sec:llm}
\input{llm}

\section{Methodology}\label{sec:methodology}
\input{methodology}

\section{Results}\label{sec:results}
\input{results}

\section{Discussion}\label{sec:discussion}
\input{discussion}

\section*{Acknowledgement}
We acknowledge the support of Groupe SIGMA.

\bibliographystyle{elsarticle-harv} 
\bibliography{main}





% \end{thebibliography}

\clearpage
\end{document}

%% file: introduction.tex
The rise of artificial intelligence (AI) and automation in various sectors, such as IT, health or even retail, invites us to question our work habits and consider greater collaboration with increasingly powerful systems to assist us with certain tasks. For example, to help with medical diagnosis or even sales prediction \citep{agossah:hal-03789529}. However, there are already many concerns about the use of these tools, particularly in the face of systems perceived as black boxes whose functioning is difficult to understand. The need for transparency is now a major issue to contribute to making AI more acceptable, given the tools for which: 1) it is difficult to understand how they work \citep{arrieta2020explainable} \citep{rudin2019stop}, 2) the certainty rate of the tool in its own behaviours and the error rate are still underutilized data \citep{agossah:hal-03789503} \citep{yin2019understanding}, and 3) the decision-making biases of these tools still question the moral and ethical considerations surrounding their design and use. There are also new concerns emerging around the use of text, code and image generation tools such as ChatGPT from OpenAI. These tools appear to be increasingly capable of understanding and generating human-like content, opening up many possibilities, but also concerns about the use of AI in the workplace.

AI can greatly benefit workers by automating simple and/or time-consuming tasks, freeing up time for more complex or higher value-added tasks. However, the strength of some AI-based tools, such as ChatGPT, which generates content, is to allow its users to access information. Like a search engine, these tools have the ability to position themselves between the requester and the information transmitter. But in this case, the response transmitted is moderated by the tool's learning from the transmitter's information. In addition, it seems important to take account individuals' perceptions and concerns when using these tools \citep{gamkrelidze2021intelligence}. One of the main challenges concerning practical cases of AI use is the need to ensure the accuracy and relaibility of these tools, in order to ensure the quality of the resultats. Furthermore, questions arise about the trust that can be placed in theses tools such as ethical issues, data security and privacy protection, or concerns about job loses due to automation \citep{agossah:hal-03789529} \citep{ferguson2019intelligence}. It is also essential to have more transparency about how these tools work to clarify the responsibility attributed to the human who cooperates with these systems.

In this article, we propose to study the perception of this new type of interaction for accessing and generating information. We therefore conduct a case study within an IT company to study employees’ perception of generative tools in terms of acceptability. After exploring what we consider to be generative tools, we will describe the methodology used and then present our results.

%% file: llm.tex
Artificial intelligence (AI) is a field that aims to design programs capable of performing tasks normally reserved for humans, and without explicit instructions. Tools for generating text, code and images are among the applications of AI that are increasingly present. These tools are based on AI models called \textit{Large Language Models} (LLMs), which are trained on huge amounts of text to learn linguistic structures and patterns. With this knowledge, LLMs are able to use advanced algorithms to automatically generate content from instructions such as keywords or phrases, creating text, code and images that appear to have been produced by humans.

For code generation, generation tools most often create computer code from natural language, comments in the code or specification. They can thus facilitate the work of developers by allowing them to generate code more quickly than in peer-programming, which avoids them writing repetitive or complex code, but the quality still seems lower than when it is exclusively produced by humans in the same given timeframe \citep{imai2022github}. In order to study the accuracy and comprehensibility of the code provided by Github Copilot, Nguyen and Nadi (2023) conducted a comparative study of Copilot's performance on four programming languages \citep{nguyen2022empirical}. The researchers gave multiple code generation requests per language to Copilot and analyses of the generated code. They conclude their study by presenting Copilot as a good starting point, capable of delivering between 60\% and 91\% of correct code depending on the language. But they did not find any significant difference between languages in terms of complexity.

To scale the performance of Copilot, Dakhel and his collaborators conducted a comparative study with human performance \citep{dakhel2022github}. The objective was also to identify whether Copilot is a good programming companion. The researchers evaluated the tool's ability to suggest correct, efficient and reproducible solutions for fundamental programming problems. In a second step, the researchers wanted to identify whether the solutions proposed by Copilot are competitive with those given by a human for the same given problems. To do this, they mobilised computer science students. The researchers concluded that the students kept a higher ratio of correct and diverse answers than the generation tool. Nevertheless, the need to repair buggy solutions and the complexity of the generated codes remain lower for Copilot. The number of studies investigating Copilot's capabilities is still relatively small at present.

Until 2022, most generative tools belonged to a particular generation category. But OpenAI's ChatGPT, which has been released to the general public since November 2022, is capable of generating text and code. It is a chatbot that now uses the GPT-4 language model to understand and interpret user queries and generate relevant responses. This tool is specialised in dialogue and can adapt to different topics and styles, but it cannot generate images by itself. ChatGPT can thus be integrated into many other fields than programming. 

We believe that generation tools offer a differentiating interaction experience, becoming a media through their ability to transmit and disseminate information. In particular, by generating content that responds to specific queries, these tools can help to communicate more effectively with target audiences and provide useful, engaging and personalised information in real time. These tools represent a major advance in the field of AI and offer new opportunities for professionals and amateurs. However, they also raise ethical and social issues related to the quality, reliability, accountability and intellectual property of the content generated. For example, the question of who is responsible for errors in the generated code, or how to deal with plagiarism issues in texts produced by LLMs.

In this study, we will focus on the perceptions of these generation tools within a department of an IT company. By examining current research and developments in the field of AI-based authoring tools, we seek to better understand how these tools are perceived by a specific professional audience. To this end, we will present our survey methodology to explore these perceptions through the lens of acceptability.

%% file: methodology.tex
\subsection{Measurement of acceptability}

The use of generative tools is becoming increasingly common in the IT field because of their potential to help generate and optimise code, create documentation, search for information, etc. We developed a customised questionnaire to explore the perception of employees of an IT company towards these tools in the context of their work. The objective of this questionnaire was to study the perception of these tools from the point of view of acceptability (intention to use). Acceptability is often presented as the sum of the representations that an individual has with regard to a technology, leading to an intention to use the latter \citep{tricot2003utility} \citep{bobillier2009adoption} \citep{pasquier2012define}. This concept is also approached from a design perspective, as the characteristics of a tool that lead to its target users intending to use it \citep{nielsen1994usability}.

To study the acceptability of generative tools, we explored some of factors influencing the acceptance or rejection of these tools (perceived ease of use, perceived usefulness, social influence, hedonic motivation) according to the litterature, using the Technology Acceptance Model (TAM) \citep{davis1989technology} \citep{atarodi2019modele} and Unified Theory of Acceptance and Use of Technology model 2 (UTAUT 2) \citep{venkatesh2012consumer}. TAM measures intention to use based on perceived usefulness and ease of use of the device, while the UTAUT 2 model focuses on more diverses determinants such as hedonic motivation and social influence. We included social influence and hedonic motivation in our study, assuming that, due to the recent emergence and widespread use of LLM-based generation tools, the social environment and attractiveness may be dominant in the intention to use. In summary, our study focuses on the acceptability of generative tools, exploring the factors that influence their adoption or rejection, using the TAM and UTAUT 2 models. We chose to explore these concepts by applying a questionnaire to the observed service.

\subsection{Hypothesis}

In this study, we have formulated several hypotheses concerning the factors that may influence the perception of generation tools within an NSE. We hypothesised that job experience, type of job, having tried a generation tool before (in a generic way), the degree of importance given to the tasks for which generation tools are used, and the frequency of use of generation tools to perform one's tasks may all have an impact on the perception of these tools. These hypotheses will be tested through the survey conducted with the employees concerned. The hypotheses are set out as follows:

\textbf{H1 :} Experience in the job can influence the perception of the tool. The more experience a person has in their job, the more informed they can be about the usefulness and effectiveness of the tool.

\textbf{H2 :} The type of job can also have an impact on the perception of the tool. Depending on the tasks and responsibilities associated with a job, the use of a generation tool may be perceived differently.

\textbf{H3 :} The frequency of use of the generation tools to carry out one's tasks can also have an impact on their perception. A person who uses these tools frequently may have a more positive view of their usefulness.

\textbf{H4 :} The degree of importance attached to the tasks for which generation tools are used can influence the perception of these tools. If a person places a lot of importance on these tasks, he or she may be more critical of the use of a tool to perform them.

%% file: results.tex
In the context of our study, we felt it was important to ask respondents about their working methods, and in particular about the strategies they use to facilitate their task when they encounter difficulties. Searching for information online and asking a colleague for help were the most common actions taken (see Figure 1).

Of the 41 people in the department, 19 responded to the survey. Of these 19 respondents, 9 use digital solutions that are not provided by the organisation to carry out their note-taking, documentation and time management tasks. Regarding code, text or image generation tools, 16 respondents already know ChatGPT and have already tried it, while 7 respondents know GitHub Copilot, of which 2 have already tried it. Finally, 6 respondents mention other generative tools such as DALL-E and Google BARD.

When asked about the frequency of use of such tools in their work, the majority of respondents stated that they never use them, while the other participants use them on an occasional basis (see Figure 2). Interestingly, 13 respondents found generative tools useful for their work, with various uses such as development support (5 respondents), optimising information retrieval (4 respondents), writing emails (3 respondents) and pre-sales, management and POC (proof of concept) advice (3 respondents). These results suggest that generative tools have potential to improve efficiency and productivity at work and can be considered acceptable by users.

\subsection{Correlation between the studied dimensions}

The correlation between the dimensions studied was assessed using Pearson's correlation. The results show that all the dimensions studied are moderately correlated with each other, with correlations ranging mainly from 0.5 to 0.846, with the exception of social influence which has the lowest correlation scores (r between .263 and .539) with all other dimensions. A correlation table is presented below:

\begin{figure}[H]
\centering
\includegraphics[width=0.70\linewidth]{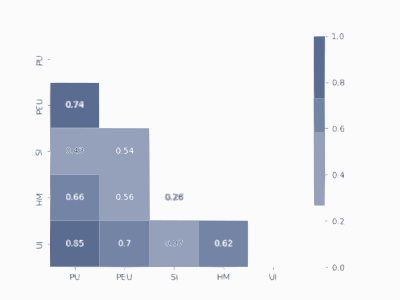}
\caption{Correlogramm of Pearson correlation between the studied dimensions of acceptability (PU: Perceived Usefulness ; PEU: Perceived Ease of Use ; SI : Social Influence ; HM : Hedonic Motivation ; UI : Usage Intention)}
\label{correlogram}
\end{figure}

These results suggest that the different dimensions studied are interconnected and therefore likely to have an impact on each other. However, social influence does not seem to be related to the other dimensions, which may indicate that social factors play a different or distinct role in the perception of LLM tools at work than the other dimensions studied.

\subsection{Perception of the generation tools according to the respondents' classification}

\begin{figure}[H]
\centering
\includegraphics[width=0.45\linewidth]{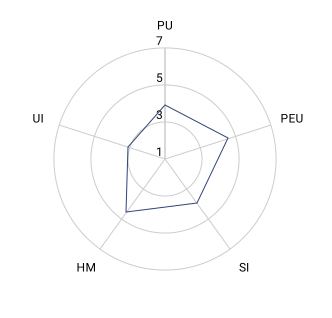}
\caption{Perception of generation tools}
\label{perceptions_globales}
\end{figure}

\begin{figure}[H]
\centering
\begin{minipage}[b]{0.45\linewidth}
    \centering
    \includegraphics[width=\textwidth]{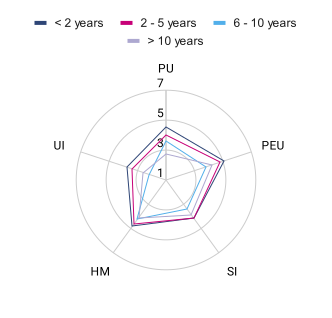}
    \caption{Perception of the generation tools according to experience in current job}
    \label{fig:exp}
\end{minipage}
\hspace{0.5cm}
\begin{minipage}[b]{0.45\linewidth}
    \centering
    \includegraphics[width=\textwidth]{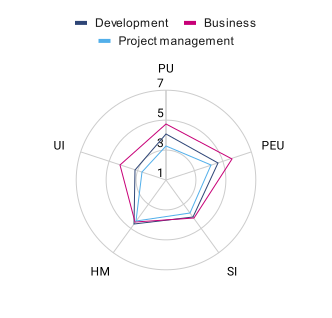}
    \caption{Perception of the generation tools according to type of job}
    \label{fig:typeofjob}
\end{minipage}
\end{figure}

\begin{figure}[H]
\centering
\begin{minipage}[b]{0.45\linewidth}
    \centering
    \includegraphics[width=\textwidth]{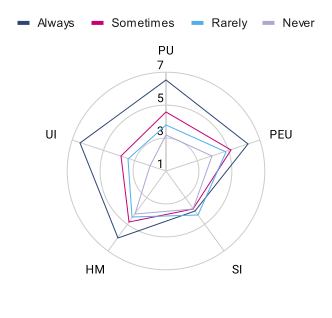}
    \caption{Perception of the generation tools according to estimate frequency of usage of generation tools in their job}
    \label{fig:frequency}
\end{minipage}
\hspace{0.5cm}
\begin{minipage}[b]{0.45\linewidth}
    \centering
    \includegraphics[width=\textwidth]{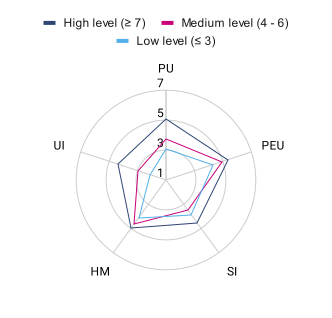}
    \caption{Perception of the generation tools according to degree of importance attached to the task}
    \label{fig:degree}
\end{minipage}
\end{figure}

In order to further explore the perception of LLM-based generation tools (see figure \ref{perceptions_globales}), we considered our sample from different perspectives. The aim was to identify whether there are differences in perceptions according to objective factors such as job experience (see figure \ref{fig:exp}), the type of job (see figure \ref{fig:typeofjob}), frequency of use of generation tools (see figure \ref{fig:frequency}) and the degree of importance attached to the task for what employees already use that kind of tools (see figure \ref{fig:degree}).

It is interesting to note certain trends in the scores presented. For example, the more experience respondents have in their jobs, the less useful they perceive generative tools to be. Or, respondents who use generative tools for high-value tasks perceive these tools as useful and easy to use. However, the social influence seems to vary very little between the different conditions. We will further analyse this in section 4.3.

\subsection{Hypothesis testing}

As previously stated, we chose to explore the perception of generative tools in the context of work is impacted by the experience in the job (\textbf{H1}), the type of job (\textbf{H2}), the frequency of use of generative tools (\textbf{H3}), the degree of importance given to the tasks for which the respondents use generative tools (\textbf{H4}) To test our hypotheses, we used the Kruskall Wallis test. This is a non-parametric statistical test, serving as an alternative to the one-factor ANOVA (see table \ref{tab:kruskall}).

\begin{table}[H]
\centering
\caption{Results of Kruskal-Wallis tests assessing the effect of different respondent classifications on the measured dimensions of acceptability (\textit{p value}) (PU: Perceived Usefulness, PEU: Perceived Ease of Use, SI: Social Influence, HM: Hedonic Motivation, UI: Usage Intention)}
\label{tab:kruskall}
\renewcommand{\arraystretch}{2}
\begin{tabular}{p{1.5cm}|p{2.3cm}p{2.3cm}p{2.3cm}p{2.3cm}}
\hline
 & \centering Work experience & \centering Job Type & \centering Estimate frequency of usage & \centering Degree of importance attached to the tasks \cr
\hline
PU & \centering 5.365 (\textit{.147}) & \centering 3.032 (\textit{.220}) & \centering 7.85 (\textit{.049}) & \centering 11.79 (\textit{.067}) \cr
PEU & \centering 1.755 (\textit{.625}) & 4.710 \centering (\textit{.095}) & \centering 5.51 (\textit{.138}) &\centering 6.84 (\textit{.336}) \cr
SI & \centering .310 (\textit{.958}) & \centering .597 (\textit{.742}) & \centering 1.00 (\textit{.801}) & \centering 4.25 (\textit{.643}) \cr
HM & \centering 3.990 (\textit{.263}) & \centering .835 (\textit{.659}) & \centering 
5.28 (\textit{.152}) & \centering 9.15 (\textit{.166}) \cr
UI & \centering 3.925 (\textit{.270}) & \centering 2.391 (\textit{.303}) & \centering 10.15 (\textit{.017}) & \centering 10.44 (\textit{.107}) \cr
\hline
\end{tabular}
\end{table}

The results of the hypothesis tests presented in the table \ref{tab:kruskall} indicate that some of our hypotheses are supported by the data, while others are not. First, regarding job experience (\textbf{H1}), none of the tests resulted in a p-value of less than .05, suggesting that there is no significant difference in the perception of generative tools based on job experience.

In terms of job type (\textbf{H2}), no p-value less than .05 was observed either, indicating that perceptions of generative tools do not differ significantly by job type.

However, the frequency of use of generative tools (\textbf{H3}) shows significant differences in perceived usefulness (H(3) = 7.85, p < .05) and intention to use (H(3)=10.15, p < .05). This suggests that respondents who use generative tools more frequently perceive them as more useful and have a higher intention to use.

Finally, for the degree of importance placed on the tasks for which respondents use generative tools (\textbf{H4}), although only one p-value is slightly greater than the threshold of .05 (perceived usefulness : H(6) = 11.79, p = .067). This may indicate a trend in which respondents who use these tools for more important tasks perceive greater usefulness. However, it is important to note that these results are not statistically significant and should therefore be interpreted with caution.

Overall, these analyses suggest that frequency of use of generative tools seems to be a key factor in understanding how respondents perceive these tools in the context of their work. Further research could focus on why frequency of use influences perceived usefulness and intention to use, as well as how to encourage wider adoption of these tools among employees.

%% file: discussion.tex
In conclusion, this study aimed to explore the perception of LLM-based generative tools in the professional context. We examined the dimensions of perceived usefulness, perceived ease of use, social influence, hedonic motivation and intention to use, and if they were impacted by objective factors such as job experience, type of job, frequency of use of generative tools and the degree of importance attached to the tasks for which these tools are used.

Our results show that generative tools are generally perceived as moderately useful and easy to use, although their adoption within the organisation studied remains limited. The dimensions studied are also moderately correlated with each other, except for social influence, which has the lowest correlation scores. Moreover, the frequency of use of the generation tools turns out to be a key factor in understanding the perception of these tools by employees. People who use them more frequently perceive these tools as more useful and have a higher intention to use them.

These findings suggest that LLM-based generation tools have potential to improve efficiency and productivity at work, but their adoption could be hindered by factors such as resistance to change, lack of awareness or lack of adequate training. Organisations seeking to encourage the use of these tools could consider initiatives to promote their adoption, such as training, workshops, or internal communication campaigns. It still conditionnate by the purpose of the implementation (improving workers agentivity, helping them for specific tasks, remplacing them, etc.). 

It is important to note that this study has some limitations, including the sample size and the fact that it focuses on a single organisation. Future research could extend this study to other work contexts and include larger samples to examine the generalisability of our findings. In addition, qualitative studies could be conducted to further our understanding of why some people adopt these tools more quickly than others, and to identify best practices for promoting their adoption.

In our perspectives, this study is the first step of a larger investigation on the acceptability of that kind of tool as a professional media. For futur work, we aim to focus on the type of requests that developers would prompt to ChatGPT specically. Investigate this point could permit us to have a better overview on the nature of the interaction with that tool for this specific public.

Ultimately, this study provides valuable insight into how employees perceive LLM-based generation tools. But it also offers resources of reflexion for organisations looking to leverage these technologies to improve the productivity and experience of their employees in getting their jobs done.